\documentclass{sigchi-ext}

\usepackage{totcount}\regtotcounter{fixmeTotalCounter}
\usepackage{datetime2} 

\newcommand{\STT}[1]{{%
\leavevmode\color{black}
\textsl{STT}}} 
\newcommand{\STTlong}[1]{{%
\leavevmode\color{black}
Scores Through-Time}}

\newcommand{\REDACT}[1]{$\Box\Box\Box$} 

\newcommand{\redactCollege}[1]{[a U.S. University]}  

\newcounter{boldifyCounter}
\newcounter{fixmeSectionCounter}
\newcounter{fixmeTotalCounter}
\makeatletter
\@addtoreset{fixmeSectionCounter}{section}
\@addtoreset{fixmeSectionCounter}{subsection}
\@addtoreset{boldifyCounter}{section}
\@addtoreset{boldifyCounter}{subsection}
\makeatother

\newcommand{\boldify}[1]{}
\ifdefined\boldifyON
	\renewcommand{\boldify}[1]{
        \par\noindent
		\stepcounter{boldifyCounter}
		\textbf{{\color{green}**}
		~\arabic{section}.\arabic{subsection}.\arabic{boldifyCounter}
		: #1} 
	}
\fi 

\newcommand{\reportOnFIXME}{%
    \newcount\iterCounter
    \iterCounter=1
    \newcount\endCounter
    \endCounter=\totvalue{fixmeTotalCounter}
    \advance \endCounter +1
    There are 
    {\color{red}\total{fixmeTotalCounter}} 
    FIX\_ME\\
    links:
    \loop
        \hyperlink{fixTag\the\iterCounter}{\#\the\iterCounter}
        \advance \iterCounter +1
    \ifnum \iterCounter < \endCounter
    \repeat
}

\newcommand{\FIXME}[1]{} 
\ifdefined\fixmeON
	\renewcommand{\FIXME}[1]{\par\noindent
		\stepcounter{fixmeSectionCounter}\stepcounter{fixmeTotalCounter}
		{\color{red}\fbox{\color{black}
			\parbox{.99\linewidth}{
				\textbf{\hypertarget{fixTag\thefixmeTotalCounter}{FIXME}	\arabic{section}.\arabic{subsection}.
        		\arabic{fixmeSectionCounter} (\color{red}
        		\#\arabic{fixmeTotalCounter}):} #1}}
        }
	}
\fi 

\newcommand{\FIXED}[1]{}
\ifdefined\fixmeON
	\renewcommand{\FIXED}[1]{\par\noindent%
		{\color{black}\fbox{\color{black}%
			\parbox{.99\columnwidth}{%
				\color{blue}#1}}%
        }
	}
\fi 

\newcommand{\draftStatus}[1]{}
\ifdefined\draftStatusON
	\renewcommand{\draftStatus}[1]{
        \hfill **#1
	}
\fi 

\def\plaintitle{%
Conceptualizing the Relationship between AI Explanations and User Agency}
\title{%
\plaintitle}

\def\plainauthor{Iyadunni Adenuga, Jonathan Dodge}

\def\plainkeywords{
    \vspace{10pt}
        Explainability; Agency; AI systems
}

\ccsdesc[500]{
    Human-centered computing~HCI theory, concepts and models
    }
    
\def\abstractText{
    \vspace{10pt}
        We grapple with the question:
        \emph{How, for whom and why should explainable artificial intelligence (XAI) aim to support the user goal of agency?}
        In particular, we analyze the relationship between agency and explanations through a user-centric lens through case studies and thought experiments.
        We find that explanation serves as one of several possible first steps for agency by allowing the user convert forethought to outcome in a more effective manner in future interactions.
        Also, we observe that XAI systems might better cater to laypersons, particularly ``tinkerers,'' when combining explanations and user control, so they can make meaningful changes.
    \vspace{12pt}
        \keywords{\plainkeywords}  
}

\numberofauthors{2}
\author{%
    \alignauthor{%
        \textbf{Iyadunni Adenuga}\\
        \affaddr{Penn State University} \\
        \affaddr{University Park, PA 16802, USA} \\
        \email{ija5027@psu.edu} 
    }\alignauthor{%
        \textbf{Jonathan Dodge}\\
        \affaddr{Penn State University} \\
        \affaddr{University Park, PA 16802, USA} \\
        \email{jxd6067@psu.edu} 
    }
    \vspace{10pt}
        \begin{abstract}
           \abstractText 
        \end{abstract}
}

\usepackage[T1]{fontenc}
\usepackage{textcomp}
\usepackage[scaled=.92]{helvet} 
\usepackage{graphicx} 
\usepackage{balance}  
\usepackage{booktabs} 
\usepackage{ragged2e} 

\usepackage{calc, enumitem} 
\usepackage{graphicx}

\CopyrightYear{2023}
\setcopyright{rightsretained}
\conferenceinfo{CHI Workshops '23}{April  23--28, 2023, Hamburg, Germany}
\isbn{978-1-4503-6819-3/20/04}
\doi{https://doi.org/10.1145/3334480.XXXXXXX}
\copyrightinfo{\acmcopyright}

\definecolor{linkColor}{RGB}{6,125,233}
\hypersetup{%
  pdftitle={\plaintitle},
  pdfauthor={\plainauthor},
  pdfkeywords={\plainkeywords},
  bookmarksnumbered,
  pdfstartview={FitH},
  colorlinks,
  citecolor=black,
  filecolor=black,
  linkcolor=black,
  urlcolor=linkColor,
  breaklinks=true,
}
\begin{document}

\maketitle







\section{Introduction}
\label{sectionIntro}

Complex technologies are commonplace in today's society, with examples including reinforcement learning, deep neural networks, or other forms of artificial intelligence (AI).
Criticisms have plagued the acceptance of these technologies due to the opaque nature of the algorithms and the erasure of user influence (i.e., creating an automated experience).
For example, high-stakes scenarios (e.g. law enforcement, medicine, etc.) traditionally require human experts that go through rigorous training, who are then accountable to human stakeholders.
Thus, it is unsurprising that such decision makers prefer worse-performing, interpretable models over opaque models~\cite{Veale2018}.

Beyond experts, laypeople also desire a level of control and understanding of the complex AI systems that affect them~\cite{woodruff2018, shneiderman2016grandchallenges}.
Legal regimes (e.g., European Union General Data Protection Regulation~\cite{GDPR} and White House Executive Order~\cite{whiteHouse2022}) align with such observations by highlighting the importance of human agency over these systems and the need for these systems to explain and justify their results.

However, making AI systems more \emph{agentic} is not as widely studied as making them \emph{explainable}.
This paper attempts to describe how designing for agency fits with XAI, namely:
1) the relationship between agency and explanations and
2) agency in scenarios with two and three user groups.

\section{What is an Explanation?
}
\label{sectionExplanation}

Explanation is a human phenomenon that strongly relates with peoples' mental models, understanding and knowledge of ``why an outcome happened''~\cite{keil2006explanation}.
Its social interactive characteristic~\cite{keil2006explanation} means there's some level of communication (which may be continuous) occurring between the explain-\emph{er} and explain-\emph{ee}.

\begin{marginfigure}
      \begin{minipage}{\marginparwidth}
        \centering
        \includegraphics [trim={4.2cm 18.7cm 4cm 0}, clip, width=\marginparwidth]{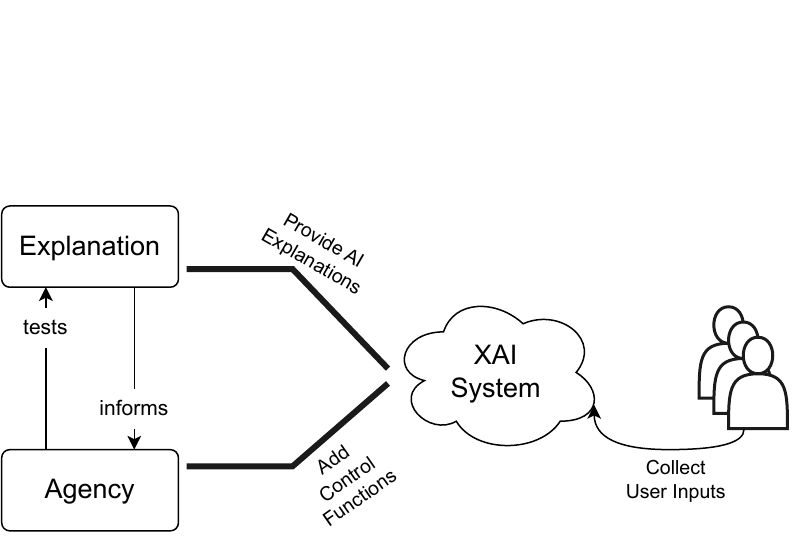}    \caption{Relationship between Agency and Explanations in an XAI system}~\label{fig:agencyExplrela}
        \vspace{30pt}
      \end{minipage}
\end{marginfigure}

Today, AI systems are a major part of our environment.
If the target users do not understand the model, they usually cannot assess or appropriately rely~\cite{schemmerIUI23} on it.
To address this, post-hoc methods aim to make opaque AI methods (e.g. neural networks, ensemble models, etc.) more ``understandable'' without compromising accuracy~\cite{arrieta2020explainable, dovsilovic2018explainable}.
There are two main approaches employed by post-hoc techniques: \emph{opaque box} (operates on the input/output boundary; e.g., LIME~\cite{ribeiro2016should}, LORE~\cite{guidotti2019factual}) and \emph{transparent box} (operates on the internal structures; e.g., deconvnet model method~\cite{zeiler2014visualizing} and network dissection~\cite{bau2017network, bau2020understanding}).




\begin{marginfigure}[90pt]
  \begin{minipage}{\marginparwidth}
    \centering
    \includegraphics [trim={0.4cm 0 0.4cm 0}, clip, width=1.05\marginparwidth]{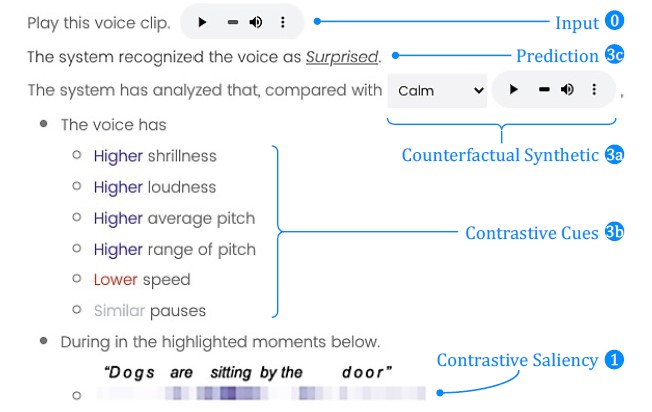}
    \caption{From Zhang and Lim paper~\protect\cite{zhang2022towards}, user interface of Counterfactual Explanation for Voice-Emotion Recognition system (best viewed digitally).}
    \label{fig:counterfactual}
  \end{minipage}
\end{marginfigure}

Existing XAI systems that utilize the opaque- and transparent-box approaches described above do not fit the requirements laid out in prior work for ``everyday'' explanations understandable to the layperson~\cite{mittelstadt2019explaining}.
AI explanations created based on human characteristics (e.g. preferences, reasoning and perception methods) are more relatable and effective~\cite{wang2019designing, zhang2022towards, lai2023selective}.
In their work about connecting existing XAI techniques to user expectations for explanations, Liao et al.~\cite{liao2020questioning} propose a ``question-driven framework'' that encourages an interactive explanation experience~\cite{liao2020questioning} via meaningful interrogative dialogue~\cite{mittelstadt2019explaining}. 
\section{What is Agency?}
\label{sectionAgency}

People have an innate need to control the course of their lives and predict the outcomes of situations, no matter the difficulty~\cite{bandura1996reflectionspart1}. 
Humans feel a sense of agency when we believe that our \emph{``conscious intention caused a voluntary action''}~\cite{wegner1999apparent}.
Agency is an internal ``human'' feeling that is outwardly expressed by intentional actions.
If people do not feel in control, they might abandon the on-going task or distrust their actions, especially in hard situations~\cite{bandura1982self}.




 
A technology that affords agency is ``flexible'' to the user's interactions inputs and interests such that they can modify their experience~\cite{wyeth2007agency, thue2010player}. 
The control a person has in a typical environment (such as while utilizing technology) can be weighted by:
1) the presence of relevant actions;
2) the relationship between the actions of a user and the outcome in the environment;
3) the ability of a user to predict the outcome of their actions, and;
4) the ability of the user to trace the cause of an outcome~\cite{thompson1998illusions}.

Researchers have shown agentic interactions have positive effects such as improved user experience and satisfaction and more appropriate trust~\cite{girgensohn2001home, huh2010incorporating, vaccaro2018illusion}. 
The many AI systems stakeholders with low technical knowledge should also experience these benefits, as per the ACM Code of Ethics: \emph{``...all people are stakeholders in computing''}~\cite{acm2022}.
 
\section{How are Agency and Explanation Related?
}
\label{sectionHowRelated}

The answer to this question is not straightforward, but we will attempt an answer for  AI systems.
Existing human-centered  XAI systems prioritize providing explanations in an understandable, visually appealing format with an assumption of improved agency in the represented artifact.
There is no \emph{direct} measure for the ``actual'' agency a user experienced while interacting with such a system.
\emph{Self-reporting} only measures agency \emph{perceived} by users, which is a proxy for ``actual'' agency. Teasing out the modalities of the relationship between explanation and agency is the first step in \textit{deducing} the ``actual'' agency in XAI systems.


The agency process starts from a person's \emph{forethought} to their performance of an action and then, observation of the action's outcome.
The aim of explanations is to improve the consumers' understanding of their environment. 
Providing explanations can contribute to consumers' sense of agency by informing their initial \emph{forethought} so  they perform the appropriate actions to successfully complete their task.
People with higher need for control are more likely to seek more information and clarifications in a work environment~\cite{goller2017human}.
This shows that even before the introduction of explanations to a scenario, an individual has an inherent agency requirement---and that such requirements will vary among users.



Studies on designing agency in AI systems, such as interactive machine learning, have primarily focused on users with technical know-how, and in its absence, requires additional technical training for end-users for them to understand and use the provided agentic functions~\cite{smithRenner2020control, dudley2018review}.
End-users with no access to technical training can still benefit from an agentic experience with explanations.


\boldify{expansion,test, interactivity, xplor, illusory}
\begin{marginfigure}[20pt]
  \begin{minipage}{\textwidth}
    \centering
    \includegraphics [width=\marginparwidth]{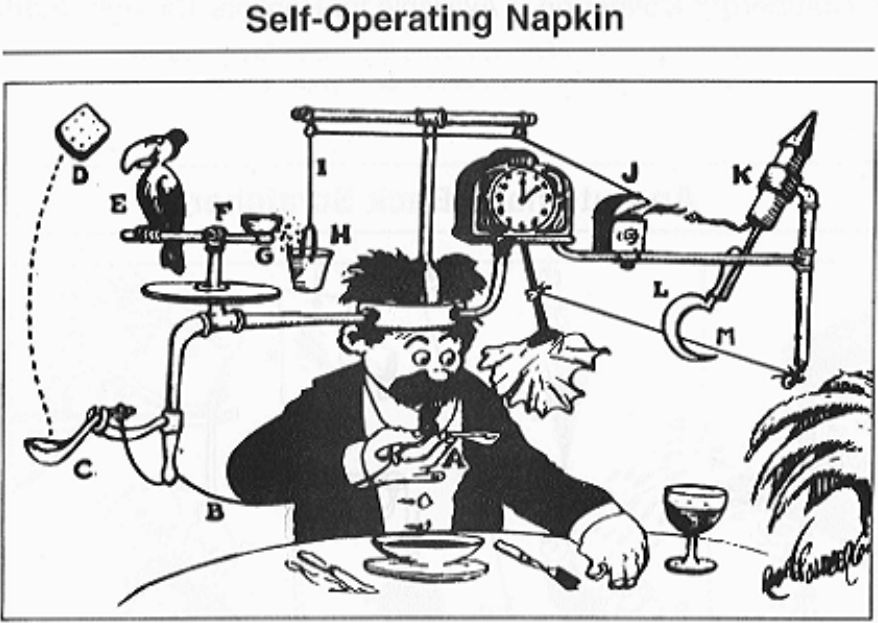}
    \caption{The original Rube Goldberg machine, as depicted in the cartoon ``Professor Butts and the Self-Operating Napkin.''
    The machine functions as follows:
    \emph{``Soup spoon (A) is raised to mouth, pulling string (B) and thereby jerking ladle (C), which throws cracker (D) past toucan (E). Toucan jumps after cracker and perch (F) tilts, upsetting seeds (G) into pail (H). Extra weight in pail pulls cord (I), which opens and ignites lighter (J), setting off skyrocket (K), which causes sickle (L) to cut string (M), allowing pendulum with attached napkin to swing back and forth, thereby wiping chin.''}
    \url{https://en.wikipedia.org/wiki/Rube_Goldberg_machine}}~\label{fig:rubeGoldberg}
  \end{minipage}
\end{marginfigure}
Users can take an active role in their absorption of an explanation.
Zhang and Lim~\cite{zhang2022towards} studied  providing relatable explanations for a vocal-emotion recognition system, the participants preferred and utilized more effectively the ``Counterfactual Sample + Cues'' explanation (Figure \ref{fig:counterfactual}).
The user interface for this explanation required active play-through and listening to alternative voices to detect vocal differences.
Another method for involving people in the explanation process is to obtain input from them to create ``selective'' explanations~\cite{lai2023selective}.
Here, the user customizes the types of received explanations to their taste.

Tastes vary, for example GenderMag~\cite{burnett2016gendermag-jrnl} identified facets describing people's cognitive styles.
One important axis is \emph{learning style}, with people who gain understanding by ``tinkering'' with the technology on one end.
To cater to tinkerers, XAI system designs should have control functions.
These functions would accept different kinds of user inputs and then provide appropriate visible outcomes, allowing the system to \emph{``be actionable''}~\cite{kulesza2015principles}.
On the other end of the learning style axis are people who gain understanding \emph{by process}.
Process-oriented learners may benefit from assessment \emph{processes}, such as  \textit{After-Action Review for AI} (AAR/AI)~\cite{dodge2021aarai}.
Later, Khanna et al.~\cite{khanna2022finding} found participants helped participants examine and effectively use explanations to identify AI faults, observing a moderate-sized practical effect.


There are situations when XAI systems cannot honor user inputs~\cite{smithRenner2020control}.
How should the system react?
For low-stakes scenarios, illusory agency may be a useful tool.
Game designers use this as a complementary mechanism to preserve their rigid game-story narrative~\cite{day2017agency, maccallum2007illusory}.
To allow for continued user agency, the user is able to observe the effect of their input but the input has no effect on the underlying algorithms.
Vaccaro et al.~\cite{vaccaro2018illusion} showed in a social media setting that users \emph{``felt more satisfied with the presence of controls''} regardless of their effectiveness.
Some everyday systems that already utilize illusory agency include crosswalk buttons and elevator close-door buttons.  
Illusory agency should only be designed to supplement the already present ``real'' agentic experience in low-stakes scenarios so as to avoid user deception and minimize the effects of ethical issues.
Example of such scenarios that might benefit from illusory agency include XAI systems in training environments~\cite{fiok2022explainable}.




\emergencystretch 3em

\section{How Does One Increase or Decrease Agency?}
\label{sectionIncreaseDecrease}

\boldify{We will use 2 examples, introduce the first Rube Goldberg machine}
We will use two examples to illustrate adjusting agency.
The first example is to consider wiping your mouth with a napkin using direct manipulation vs with a Rube Goldberg machine, depicted in Figure~\ref{fig:rubeGoldberg}.
Rube Goldberg machines are famous for having a simple input, which then initiates a complicated chain reaction  generating a simple output.
In this case, the simple input is lifting a spoon, the long chain reaction is via the crackers, toucan, string, etc., and the simple output is wiping one's mouth.

\boldify{And here is enough information about these machines that if you have never heard of them that you can follow along}


\begin{marginfigure}[10pt]
      \begin{minipage}{\marginparwidth}
        \centering
        \includegraphics
        [width=\marginparwidth]{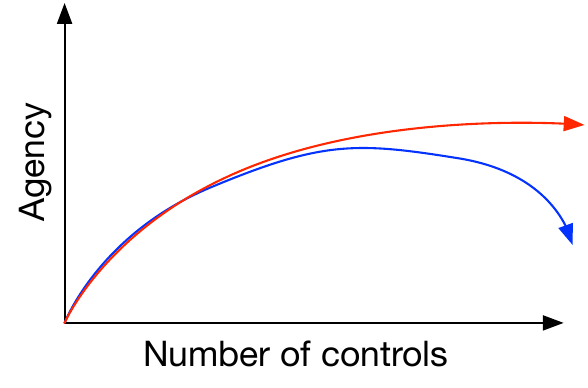}    
        \caption{%
        Notional curve depicting agency as a function of number of controls.
        When moving from no controls to few controls, agency gain enjoys a direct relationship.
        But at some point, the extra controls will overwhelm the user, either taking the form of a plateau (red curve) or even a downturn (blue curve).
        }~\label{fig:buttonCount}
        \vspace{30pt}
      \end{minipage}
\end{marginfigure}

\begin{margintable}[-35pt]
    \fbox{
    \begin{minipage}{0.925\marginparwidth}
    \centering
    \begin{tabular}{@{}p{\columnwidth}@{}}
    
    1)~Add/remove source documents
    \\\hline
    2)~Add/remove sections, where sections are subtopics of the document title
    \\\hline
    3)~Add/remove words and/or sentences
    \\\hline
    4)~View the order relation of the summary sentences in a concept graph
    \\\hline
    5)~View actual sentences contributed by each document to the overall summary output
    \\\hline
    \end{tabular}
    \caption{Key functionality found in Living Documents.}
    \label{LDfunctionality}
    \end{minipage}
    }
\end{margintable}


As Figure~\ref{fig:rubeGoldberg} shows, the machine automates the functioning of the napkin to the point that its use is involuntary.
Suppose we changed the simple input to be pushing a button, which is more typical of modern technology.
Now, consider how much agency the user has in  each case.
It seems fairly obvious that agency would be highest with direct manipulation, then with button-interface Rube Goldberg machine, and finally the unmodified Rube Goldberg machine.
The reasoning is that with direct manipulation, one could manifest whatever wiping approach they desire: arbitrary direction, length of time, and so on.
Note that all of the previously compared interactions lead to same \emph{outcome}, but are different in terms of controllability~\cite{roy2019controllability}.
According to Shneiderman~\cite{shneiderman2020human}, high levels of automation and human control can co-exist in a technical artifact.
They illustrate this in their description of the digital camera and elevator where agency is afforded by the inclusion of a button and settings page respectively.
This is similar to the surface-level agency button introduced above, to the Rube Goldberg machine.
Would increasing agency require addition of extra buttons/settings or introducing a more manual and influential process (i.e., less automation)?
If we assume adding a feature and accompanying button increases perceived agency, what is the amount of UI complexity at which agency gains diminish or turn negative (See Figure~\ref{fig:buttonCount} for an illustration)?


\boldify{short paragraph here to keep emphasis on these two key points: 1. tension exists, 2. measures are lacking}

Thus, we have illustrated a tension between manual processes (which have the highest agency) vs automation (which reduces agency).
The open question is, how much agency does one lose when a process undergoes automation?
\boldify{Now to tell you about the second example, which has a measurement strategy attendant}
To answer this question, we turn to our second example: Living Documents~\cite{adenuga2022living}.
Living Documents is an interactive multi-document text summarization system, providing the control functions found in Table~\ref{LDfunctionality}.

\boldify{Here are the high/med/low agency treatments we have devised for this domain}

 What would agency treatment levels look like in Living Documents?
 The highest-level of agency is full access to all functionality in the system (full-agency, see Table~\ref{LDfunctionality})  while the automation level has no user controls (no-agency).
 This means it would work like a typical text summarization system such that the only input-output operation is the user providing the source documents and receiving the summary result.
 The interaction designer can decide on intermediate level(s) based on specified criteria.
 Our criterion is ``magnitude of impact'' (document $\rightarrow$ sections $\rightarrow$ sentences/words), so our some-agency treatment has functionality 3 to 5 in Table~\ref{LDfunctionality}.
 

\boldify{and now we get to the measurement strategy}

\section{Agency and/or Explanations, for whom?}
\label{sectionForWhom}

\boldify{prepare the reader for a big switch}
Now, we would like to broaden the previously discussed system-user agency cases to where there are \emph{multiple types} of users, leading to a more complex tradeoff relationship.
From an explanation perspective, we know that explanations may need to account for different domain expertise, cognitive abilities, context of use, and audience.
Users have varied needs for an XAI system~\cite{liao2020questioning} and do not have a homogeneous process for interacting with models~\cite{smithRenner2020control}.
The agency perspective is less well studied.

\boldify{introduce the case study}
Consider the case of a rideshare application called Co-opRide, which is an algorithmic manager for \emph{two} user groups: drivers and riders.
What is the right agency balance to strike between these three parties (the third is Co-opRide)?
Suppose Co-opRide offers a design feature that drivers can veto riders.
This would increase the agency of the drivers at the expense of the riders' agency, as well as that of the system provider.
Imagine being a rider receiving vetoes from several drivers, based on your low population density location or even worse,  cultural markers present in your name.
This might lead to long wait times and negative customer sentiment.
Should the XAI system alert riders that a driver vetoed them?
Each time?
How should the XAI system provide notification of the veto and/or explain the decision?
If there are no satisfying agentic actions available, why should the system provide explanation at all?
In the case where the rider's waiting time increases over time as a result of receiving multiple driver vetoes, provision of explanations by the algorithmic XAI platform becomes even more imperative.  


\boldify{Move into game theory framing of this}
The example of the driver veto feature suggests ``The Agency Tradeoff Game'' might be zero-sum, though this it is not totally clear that it cannot be positive-sum or negative-sum.
It is also an open question whether or not a stable solution exists. 
As an example, ``Hotelling's game'' (see margin) has a stable solution with two players, but the three-player version has no stable solution (\cite{osborne2009gametheory}, Chapter 3).

\marginpar{%
    \vspace{-183pt} \fbox{%
    \begin{minipage}{0.975\marginparwidth}
        \textbf{\underline{Hotelling's Game}~\cite{Hotelling1929StabilityinCompetition}}
        
        Suppose two competing shops are located along the length of street, with customers spread equally along the street.
        Each customer will always choose the nearer shop.\\

        With two shops, the consumer ideal has shops at $\frac{1}4$ and $\frac{3}4$.
        However, this is unstable,
        since both shops can claim more customers by moving toward the middle.
        The stable solution has both at $\frac{1}2$.
        (E.g., this is why Lowe's and Home Depot are often co-located.)
    \end{minipage}}\label{hotelling} 
}

\boldify{social explanations, collective agency}
XAI platforms interact with groups of humans; as a result, agency occurring on a \emph{collective} basis becomes relevant.
When individuals perform a joint action, they can feel individual and/or collective (joint) agency~\cite{loehr2022sense}.
The individual perceived self-efficacy of multiple members of a group forms collective efficacy which can lead to meaningful social change~\cite{bandura1982self, bandura1996reflections}.
People are usually interested in the experiences of their fellow people and this has led to calls for social explanations~\cite{liao2020questioning, ehsan2021expanding}.
Enabling social explanations means there can be a joint platform for ``knowledge sharing'' and ``social learning''~\cite{ehsan2021expanding}. People can confidently contest the decisions by an AI system and if some form of collective agency has been designed in the system, they can effect popular meaningful change.


For an example of collective agency as a result of the conditions in an AI system, consider work by Calacci and Pentland~\cite{Calacci2022worker}.
Shipt, a grocery delivery service with an AI algorithmic manager, was initially explainable and transparent about its wage calculation process.
The introduction of a wage-processing opaque-box algorithm to Shipt led to the implementation of social explanations, albeit on a platform (called Shipt Calculator~\cite{Calacci2022worker}) external to the AI system.
Workers anonymously provided screenshots of their payment history and the Shipt Calculator aggregated the payment information and provided the observed wage difference to the workers.
These authors discovered a paycut for 41 percent of the workers in their study.
Similar occurrences with Doordash led to change in the pay and tipping model as well as a class-action lawsuit~\cite{newman_2019, keck_2019}. 
XAI systems that allow social explanations and collective agency would ensure a cooperative approach so that only beneficial improvements are implemented on the platform.

\FIXME{JED@IJA: Write around 1 paragraph about Dan Calacci's work on wage transparency for delivery drivers as a concrete example of this kind of social change, knowledge sharing, etc.//
IJA@JED: Done-ish\\
JED@IJA: new paragraph looks good. Copy edit to a whatever degree you would like and submit when ready!\\
}

\section{Concluding Remarks}
\label{sectionConclusion}

We do not claim that our statement of the relationship between agency and explanation is complete.
As an example, perhaps agency and explanations might share multiple simultaneous relationships.
We believe that recognizing and formalizing these relationship(s) would ensure that XAI designers take the closely-related extra step of designing for \emph{agency} while working on explainability. 




\bibliographystyle{SIGCHI-Reference-Format}
\bibliography{00-paper,00-paper1}


\begin{thebibliography}{00}


\ifx \showCODEN    \undefined \def \showCODEN     #1{\unskip}     \fi
\ifx \showDOI      \undefined \def \showDOI       #1{{\tt DOI:}\penalty0{#1}\ }
  \fi
\ifx \showISBNx    \undefined \def \showISBNx     #1{\unskip}     \fi
\ifx \showISBNxiii \undefined \def \showISBNxiii  #1{\unskip}     \fi
\ifx \showISSN     \undefined \def \showISSN      #1{\unskip}     \fi
\ifx \showLCCN     \undefined \def \showLCCN      #1{\unskip}     \fi
\ifx \shownote     \undefined \def \shownote      #1{#1}          \fi
\ifx \showarticletitle \undefined \def \showarticletitle #1{#1}   \fi
\ifx \showURL      \undefined \def \showURL       #1{#1}          \fi

\bibitem{acm2022}
{{ACM}}. 2018.
\newblock ACM Code of Ethics and Professional Conduct.
\newblock   (2018).
\newblock
\showURL{%
\url{https://www.acm.org/code-of-ethics}}


\bibitem{adenuga2022living}
{Iyadunni~J Adenuga}, {Benjamin~V Hanrahan}, {Chen Wu}, {and} {Prasenjit
  Mitra}. 2022.
\newblock \showarticletitle{Living Documents: Designing for User Agency over
  Automated Text Summarization}. In {\em CHI Conference on Human Factors in
  Computing Systems Extended Abstracts}. 1--6.
\newblock


\bibitem{arrieta2020explainable}
{Alejandro~Barredo Arrieta}, {Natalia D{\'\i}az-Rodr{\'\i}guez}, {Javier
  Del~Ser}, {Adrien Bennetot}, {Siham Tabik}, {Alberto Barbado}, {Salvador
  Garc{\'\i}a}, {Sergio Gil-L{\'o}pez}, {Daniel Molina}, {Richard Benjamins},
  {and} {others}. 2020.
\newblock \showarticletitle{Explainable Artificial Intelligence (XAI):
  Concepts, taxonomies, opportunities and challenges toward responsible AI}.
\newblock {\em Information fusion\/}  {58} (2020), 82--115.
\newblock


\bibitem{bandura1982self}
{Albert Bandura}. 1982.
\newblock \showarticletitle{Self-efficacy mechanism in human agency.}
\newblock {\em American psychologist\/} {37}, 2 (1982), 122.
\newblock


\bibitem{bandura1996reflectionspart1}
{Albert Bandura}. 1996a.
\newblock \showarticletitle{Reflections on human agency: Part I}.
\newblock {\em Constructivism in the Human Sciences\/} {1}, 2 (1996), 3.
\newblock


\bibitem{bandura1996reflections}
{Albert Bandura}. 1996b.
\newblock \showarticletitle{Reflections on human agency: Part II}.
\newblock {\em Constructivism in the Human Sciences\/} {1}, 3/4 (1996), 5.
\newblock


\bibitem{bau2017network}
{David Bau}, {Bolei Zhou}, {Aditya Khosla}, {Aude Oliva}, {and} {Antonio
  Torralba}. 2017.
\newblock \showarticletitle{Network dissection: Quantifying interpretability of
  deep visual representations}. In {\em Proceedings of the IEEE conference on
  computer vision and pattern recognition}. 6541--6549.
\newblock


\bibitem{bau2020understanding}
{David Bau}, {Jun-Yan Zhu}, {Hendrik Strobelt}, {Agata Lapedriza}, {Bolei
  Zhou}, {and} {Antonio Torralba}. 2020.
\newblock \showarticletitle{Understanding the role of individual units in a
  deep neural network}.
\newblock {\em Proceedings of the National Academy of Sciences\/} {117}, 48
  (2020), 30071--30078.
\newblock


\bibitem{burnett2016gendermag-jrnl}
{Margaret Burnett}, {Simone Stumpf}, {Jamie Macbeth}, {Stephann Makri}, {Laura
  Beckwith}, {Irwin Kwan}, {Anicia Peters}, {and} {William Jernigan}. 2016.
\newblock \showarticletitle{GenderMag: A method for evaluating software's
  gender inclusiveness}.
\newblock {\em Interacting with Computers\/} {28}, 6 (2016), 760--787.
\newblock


\bibitem{Calacci2022worker}
{Dan Calacci} {and} {Alex Pentland}. 2022.
\newblock \showarticletitle{Bargaining with the Black-Box: Designing and
  Deploying Worker-Centric Tools to Audit Algorithmic Management}.
\newblock {\em Proc. ACM Hum.-Comput. Interact.\/} {6}, CSCW2, Article 428 (nov
  2022), 24 pages.
\newblock
\showDOI{%
\url{http://dx.doi.org/10.1145/3570601}}


\bibitem{day2017agency}
{Timothy Day} {and} {Jichen Zhu}. 2017.
\newblock \showarticletitle{Agency informing techniques: Communicating player
  agency in interactive narratives}. In {\em Proceedings of the 12th
  International Conference on the Foundations of Digital Games}. 1--4.
\newblock


\bibitem{dodge2021aarai}
{Jonathan Dodge}, {Roli Khanna}, {Jed Irvine}, {Kin-ho Lam}, {Theresa Mai},
  {Zhengxian Lin}, {Nicholas Kiddle}, {Evan Newman}, {Andrew Anderson}, {Sai
  Raja}, {Caleb Matthews}, {Christopher Perdriau}, {Margaret Burnett}, {and}
  {Alan Fern}. 2021.
\newblock \showarticletitle{After-Action Review for AI (AAR/AI)}.
\newblock {\em ACM Trans. Interact. Intell. Syst.\/} {11}, 3–4, Article 29
  (sep 2021), 35 pages.
\newblock
\showISSN{2160-6455}
\showDOI{%
\url{http://dx.doi.org/10.1145/3453173}}


\bibitem{dovsilovic2018explainable}
{Filip~Karlo Do{\v{s}}ilovi{\'c}}, {Mario Br{\v{c}}i{\'c}}, {and} {Nikica
  Hlupi{\'c}}. 2018.
\newblock \showarticletitle{Explainable artificial intelligence: A survey}. In
  {\em 2018 41st International convention on information and communication
  technology, electronics and microelectronics (MIPRO)}. IEEE, 0210--0215.
\newblock


\bibitem{dudley2018review}
{John~J Dudley} {and} {Per~Ola Kristensson}. 2018.
\newblock \showarticletitle{A review of user interface design for interactive
  machine learning}.
\newblock {\em ACM Transactions on Interactive Intelligent Systems (TiiS)\/}
  {8}, 2 (2018), 1--37.
\newblock


\bibitem{ehsan2021expanding}
{Upol Ehsan}, {Q~Vera Liao}, {Michael Muller}, {Mark~O Riedl}, {and} {Justin~D
  Weisz}. 2021.
\newblock \showarticletitle{Expanding explainability: Towards social
  transparency in ai systems}. In {\em Proceedings of the 2021 CHI Conference
  on Human Factors in Computing Systems}. 1--19.
\newblock


\bibitem{fiok2022explainable}
{Krzysztof Fiok}, {Farzad~V Farahani}, {Waldemar Karwowski}, {and} {Tareq
  Ahram}. 2022.
\newblock \showarticletitle{Explainable artificial intelligence for education
  and training}.
\newblock {\em The Journal of Defense Modeling and Simulation\/} {19}, 2
  (2022), 133--144.
\newblock


\bibitem{GDPR}
{GDPR}. 2018.
\newblock European Union General Data Protection Regulation, Article 15 -
  ``Right of access by the data subject''.
\newblock   (2018).
\newblock
\showURL{%
\url{http://www.privacy-regulation.eu/en/article-15-right-of-access-by-the-data-subject-GDPR.htm}}
\newblock
\shownote{Accessed: 1/16/2019.}


\bibitem{girgensohn2001home}
{Andreas Girgensohn}, {Sara~A Bly}, {Frank Shipman}, {John~S Boreczky}, {and}
  {Lynn Wilcox}. 2001.
\newblock \showarticletitle{Home Video Editing Made Easy-Balancing Automation
  and User Control.}. In {\em INTERACT}, Vol.~1. 464--471.
\newblock


\bibitem{goller2017human}
{Michael Goller} {and} {Christian Harteis}. 2017.
\newblock \showarticletitle{Human agency at work: Towards a clarification and
  operationalisation of the concept}.
\newblock In {\em Agency at work}. Springer, 85--103.
\newblock


\bibitem{guidotti2019factual}
{Riccardo Guidotti}, {Anna Monreale}, {Fosca Giannotti}, {Dino Pedreschi},
  {Salvatore Ruggieri}, {and} {Franco Turini}. 2019.
\newblock \showarticletitle{Factual and counterfactual explanations for black
  box decision making}.
\newblock {\em IEEE Intelligent Systems\/} {34}, 6 (2019), 14--23.
\newblock


\bibitem{Hotelling1929StabilityinCompetition}
{Harold Hotelling}. 1929.
\newblock \showarticletitle{Stability in Competition}.
\newblock {\em The Economic Journal\/} {39}, 153 (1929), 41--57.
\newblock
\showISSN{00130133}
\showURL{%
\url{http://www.jstor.org/stable/2224214}}


\bibitem{whiteHouse2022}
{White House}. 2022.
\newblock \showarticletitle{Blueprint for an AI Bill of Rights}.
\newblock  (2022).
\newblock
\showURL{%
\url{https://www.whitehouse.gov/ostp/ai-bill-of-rights/}}
\newblock
\shownote{Last accessed: 10/13/22.}


\bibitem{huh2010incorporating}
{Jina Huh}, {Martha Pollack}, {Hadi Katebi}, {Karem Sakallah}, {and} {Ned
  Kirsch}. 2010.
\newblock \showarticletitle{Incorporating user control in automated interactive
  scheduling systems}. In {\em Proceedings of the 8th ACM Conference on
  Designing Interactive Systems}. 306--309.
\newblock


\bibitem{keck_2019}
{Catie Keck}. 2019.
\newblock DoorDash tip-skimming scheme prompts class action lawsuit seeking all
  those tips that didn't go to drivers.
\newblock   (Jul 2019).
\newblock
\showURL{%
\url{https://gizmodo.com/doordash-tip-skimming-scheme-prompts-clash-action-lawsu-1836820630}}


\bibitem{keil2006explanation}
{Frank~C Keil}. 2006.
\newblock \showarticletitle{Explanation and understanding}.
\newblock {\em Annual review of psychology\/}  {57} (2006), 227.
\newblock


\bibitem{khanna2022finding}
{Roli Khanna}, {Jonathan Dodge}, {Andrew Anderson}, {Rupika Dikkala}, {Jed
  Irvine}, {Zeyad Shureih}, {Kin-ho Lam}, {Caleb~R Matthews}, {Zhengxian Lin},
  {Minsuk Kahng}, {and} {others}. 2022.
\newblock \showarticletitle{Finding AI’s faults with AAR/AI: An empirical
  study}.
\newblock {\em ACM Transactions on Interactive Intelligent Systems (TiiS)\/}
  {12}, 1 (2022), 1--33.
\newblock


\bibitem{kulesza2015principles}
{Todd Kulesza}, {Margaret Burnett}, {Weng-Keen Wong}, {and} {Simone Stumpf}.
  2015.
\newblock \showarticletitle{Principles of explanatory debugging to personalize
  interactive machine learning}. In {\em Proceedings of the 20th international
  conference on intelligent user interfaces}. 126--137.
\newblock


\bibitem{lai2023selective}
{Vivian Lai}, {Yiming Zhang}, {Chacha Chen}, {Q~Vera Liao}, {and} {Chenhao
  Tan}. 2023.
\newblock \showarticletitle{Selective Explanations: Leveraging Human Input to
  Align Explainable AI}.
\newblock {\em arXiv preprint arXiv:2301.09656\/} (2023).
\newblock


\bibitem{liao2020questioning}
{Q~Vera Liao}, {Daniel Gruen}, {and} {Sarah Miller}. 2020.
\newblock \showarticletitle{Questioning the AI: informing design practices for
  explainable AI user experiences}. In {\em Proceedings of the 2020 CHI
  Conference on Human Factors in Computing Systems}. 1--15.
\newblock


\bibitem{loehr2022sense}
{Janeen~D Loehr}. 2022.
\newblock \showarticletitle{The sense of agency in joint action: An integrative
  review}.
\newblock {\em Psychonomic Bulletin \& Review\/} (2022), 1--29.
\newblock


\bibitem{maccallum2007illusory}
{Esther MacCallum-Stewart} {and} {Justin Parsler}. 2007.
\newblock \showarticletitle{Illusory agency in vampire: The
  masquerade--Bloodlines}.
\newblock {\em Dichtung Digital. Journal f{\"u}r Kunst und Kultur digitaler
  Medien\/} {9}, 1 (2007), 1--17.
\newblock


\bibitem{mittelstadt2019explaining}
{Brent Mittelstadt}, {Chris Russell}, {and} {Sandra Wachter}. 2019.
\newblock \showarticletitle{Explaining explanations in AI}. In {\em Proceedings
  of the conference on fairness, accountability, and transparency}. 279--288.
\newblock


\bibitem{newman_2019}
{Andy Newman}. 2019.
\newblock Doordash changes tipping model after uproar from customers.
\newblock   (Jul 2019).
\newblock
\showURL{%
\url{https://www.nytimes.com/2019/07/24/nyregion/doordash-tip-policy.html?action=click&amp};
\url{module=Intentional&amp};
\url{pgtype=Article}}


\bibitem{osborne2009gametheory}
{{Martin J.} Osborne}. 2004.
\newblock {\em An introduction to game theory}.
\newblock Oxford Univ. Press, New York, NY [u.a.].
\newblock
\showISBNx{978-0-19-512895-6}
\showURL{%
\url{http://gso.gbv.de/DB=2.1/CMD?ACT=SRCHA&SRT=YOP&IKT=1016&TRM=ppn+369342747&sourceid=fbw_bibsonomy}}


\bibitem{ribeiro2016should}
{Marco~Tulio Ribeiro}, {Sameer Singh}, {and} {Carlos Guestrin}. 2016.
\newblock \showarticletitle{" Why should i trust you?" Explaining the
  predictions of any classifier}. In {\em Proceedings of the 22nd ACM SIGKDD
  international conference on knowledge discovery and data mining}. 1135--1144.
\newblock


\bibitem{roy2019controllability}
{Quentin Roy}, {Futian Zhang}, {and} {Daniel Vogel}. 2019.
\newblock \showarticletitle{Automation Accuracy Is Good, but High
  Controllability May Be Better}. In {\em Proceedings of the 2019 CHI
  Conference on Human Factors in Computing Systems} {\em (CHI '19)}.
  Association for Computing Machinery, New York, NY, USA, 1–8.
\newblock
\showISBNx{9781450359702}
\showDOI{%
\url{http://dx.doi.org/10.1145/3290605.3300750}}


\bibitem{schemmerIUI23}
{Max Schemmer}, {Niklas Kuehl}, {Carina Benz}, {Andrea Bartos}, {and} {Gerhard
  Satzger}. 2023.
\newblock \showarticletitle{Appropriate Reliance on AI Advice:
  Conceptualization and the Effect of Explanations}. In {\em Proceedings of the
  28th International Conference on Intelligent User Interfaces} {\em (IUI
  '23)}. Association for Computing Machinery, New York, NY, USA, 410–422.
\newblock
\showISBNx{9798400701061}
\showDOI{%
\url{http://dx.doi.org/10.1145/3581641.3584066}}


\bibitem{shneiderman2020human}
{Ben Shneiderman}. 2020.
\newblock \showarticletitle{Human-centered artificial intelligence: Reliable,
  safe \& trustworthy}.
\newblock {\em International Journal of Human--Computer Interaction\/} {36}, 6
  (2020), 495--504.
\newblock


\bibitem{shneiderman2016grandchallenges}
{Ben Shneiderman}, {Catherine Plaisant}, {Maxine Cohen}, {Steven Jacobs},
  {Niklas Elmqvist}, {and} {Nicholoas Diakopoulos}. 2016.
\newblock \showarticletitle{Grand Challenges for HCI Researchers}.
\newblock {\em Interactions\/} {23}, 5 (aug 2016), 24–25.
\newblock
\showISSN{1072-5520}
\showDOI{%
\url{http://dx.doi.org/10.1145/2977645}}


\bibitem{smithRenner2020control}
{Alison Smith-Renner}, {Varun Kumar}, {Jordan Boyd-Graber}, {Kevin Seppi},
  {and} {Leah Findlater}. 2020.
\newblock \showarticletitle{Digging into User Control: Perceptions of Adherence
  and Instability in Transparent Models}. In {\em Proceedings of the 25th
  International Conference on Intelligent User Interfaces} {\em (IUI ’20)}.
  Association for Computing Machinery, New York, NY, USA, 519–530.
\newblock
\showISBNx{9781450371186}
\showDOI{%
\url{http://dx.doi.org/10.1145/3377325.3377491}}


\bibitem{thompson1998illusions}
{Suzanne~C Thompson}, {Wade Armstrong}, {and} {Craig Thomas}. 1998.
\newblock \showarticletitle{Illusions of control, underestimations, and
  accuracy: a control heuristic explanation.}
\newblock {\em Psychological bulletin\/} {123}, 2 (1998), 143.
\newblock


\bibitem{thue2010player}
{David Thue}, {Vadim Bulitko}, {Marcia Spetch}, {and} {Trevon Romanuik}. 2010.
\newblock \showarticletitle{Player agency and the relevance of decisions}. In
  {\em Joint International Conference on Interactive Digital Storytelling}.
  Springer, 210--215.
\newblock


\bibitem{vaccaro2018illusion}
{Kristen Vaccaro}, {Dylan Huang}, {Motahhare Eslami}, {Christian Sandvig},
  {Kevin Hamilton}, {and} {Karrie Karahalios}. 2018.
\newblock \showarticletitle{The Illusion of Control: Placebo Effects of Control
  Settings}. In {\em Proceedings of the 2018 CHI Conference on Human Factors in
  Computing Systems} {\em (CHI ’18)}. Association for Computing Machinery,
  New York, NY, USA, 1–13.
\newblock
\showISBNx{9781450356206}
\showDOI{%
\url{http://dx.doi.org/10.1145/3173574.3173590}}


\bibitem{Veale2018}
{Michael Veale}, {Max Van~Kleek}, {and} {Reuben Binns}. 2018.
\newblock \showarticletitle{Fairness and Accountability Design Needs for
  Algorithmic Support in High-Stakes Public Sector Decision-Making}. In {\em
  Proceedings of the 2018 CHI Conference on Human Factors in Computing Systems}
  {\em (CHI '18)}. ACM, New York, NY, USA, Article 440, 14 pages.
\newblock
\showISBNx{978-1-4503-5620-6}
\showDOI{%
\url{http://dx.doi.org/10.1145/3173574.3174014}}


\bibitem{wang2019designing}
{Danding Wang}, {Qian Yang}, {Ashraf Abdul}, {and} {Brian~Y Lim}. 2019.
\newblock \showarticletitle{Designing theory-driven user-centric explainable
  AI}. In {\em Proceedings of the 2019 CHI conference on human factors in
  computing systems}. 1--15.
\newblock


\bibitem{wegner1999apparent}
{Daniel~M Wegner} {and} {Thalia Wheatley}. 1999.
\newblock \showarticletitle{Apparent mental causation: Sources of the
  experience of will.}
\newblock {\em American psychologist\/} {54}, 7 (1999), 480.
\newblock


\bibitem{woodruff2018}
{Allison Woodruff}, {Sarah~E. Fox}, {Steven Rousso-Schindler}, {and} {Jeffrey
  Warshaw}. 2018.
\newblock \showarticletitle{A Qualitative Exploration of Perceptions of
  Algorithmic Fairness}. In {\em Proceedings of the 2018 CHI Conference on
  Human Factors in Computing Systems} {\em (CHI '18)}. Association for
  Computing Machinery, New York, NY, USA, 1–14.
\newblock
\showISBNx{9781450356206}
\showDOI{%
\url{http://dx.doi.org/10.1145/3173574.3174230}}


\bibitem{wyeth2007agency}
{Peta Wyeth}. 2007.
\newblock \showarticletitle{Agency, tangible technology and young children}. In
  {\em Proceedings of the 6th international conference on Interaction design
  and children}. 101--104.
\newblock


\bibitem{zeiler2014visualizing}
{Matthew~D Zeiler} {and} {Rob Fergus}. 2014.
\newblock \showarticletitle{Visualizing and understanding convolutional
  networks}. In {\em European conference on computer vision}. Springer,
  818--833.
\newblock


\bibitem{zhang2022towards}
{Wencan Zhang} {and} {Brian~Y Lim}. 2022.
\newblock \showarticletitle{Towards relatable explainable AI with the
  perceptual process}. In {\em Proceedings of the 2022 CHI Conference on Human
  Factors in Computing Systems}. 1--24.
\newblock


\end{thebibliography}

\end{document}